# Altered Dielectric Behaviour, Structure and Dynamics of Nanoconfined Dipolar Liquids: Signatures of Enhanced Cooperativity

Sayantan Mondal, Subhajit Acharya, and Biman Bagchi*

*Solid State and Structural Chemistry Unit, Indian Institute of Science, Bangalore, India*

*corresponding author email: bbagchi@iisc.ac.in

Spherical confinement can alter the properties of a dipolar fluid in several different ways. In an atomistic molecular dynamics simulation study of two different dipolar liquids (SPC/E water and a model Stockmayer fluid) confined to nanocavities of different radii ranging from $R_c$=1nm to 4nm, we find that the Kirkwood correlation factor remains surprisingly small in water, but not so in model Stockmayer liquid. This gives rise to an anomalous ultrafast relaxation of the total dipole moment time correlation function (DMTCF). The static dielectric constant of water under nanoconfinement (computed by employing Clausius-Mossotti equation, the only exact relation) exhibits a strong dependence on the size of the nanocavity with a remarkably slow convergence to the bulk value. Interestingly, the value of the volume becomes ambiguous in this nanoworld. It is determined by the liquid-surface interaction potential and is to be treated with care because of the sensitivity of the Clausius-Mossotti equation to the volume of the nanosphere. We discover that the DMTCF for confined water exhibit a bimodal *1/f* noise power spectrum. We also comment on the applicability of certain theoretical formalisms that become dubious in the nanoworld.

## I. Introduction

Spherical confinement can alter the properties of a fluid in several different ways. The effects can be particularly novel in dipolar liquids which exhibit long range orientational correlations. Surface induced changes in the fluid can propagate inside and interfere with the same from the opposite directions.[1] Thus, one can anticipate a possible synergy between surface effects and confinement. The situation can be particularly intricate for liquid water because of its extensive hydrogen bond network, and also large dielectric constant.

Confined water is omnipresent in nature, found in porous materials, aerosols, reverse micelles, within biological cells, and also at the surfaces (or inside the cavities) of macromolecules. These water molecules are deeply influenced by water-surface interactions which alter the structure, dynamics, and chemical reaction kinetics of solvated/confined species.[2-5] The study of solvation and charge transfer processes in dipolar liquids has become an intensely active area of research in the past few decades.[6-13] In recent years, nanoconfined fluids have received enormous attention because of the emergence of several unanticipated structural and dynamical properties.[14-22]

Under confinement, noticeable modulations occur in the phase behaviour, ion transport, reaction pathways, and chemical equilibrium of liquids. Water seems to exhibit enhanced self-dissociation under confinement.[16] This increases the ionic product and affects other physicochemical properties. Electrospray experiments show a marked increase in the reaction rate and yield in aqueous droplet medium.[20, 23-24] Some reactions adopt different mechanisms that lead to unexpected products.[25] This is partly because of the increased encounter probability among reactants. Experiments and theoretical investigations advocate the emergence of both faster and slower (than the bulk) relaxation timescales in confined water.[1, 26-28] This is a trademark of dynamical heterogeneity. Interestingly, an Ising model-based study explains the *faster than bulk* relaxation in terms of propagating destructive interference among orientational correlations from opposite surfaces, and the slower relaxation of water close to the surface.[1]

As initial discussion of processes such as solvation and electron transfer reactions invoke a continuum model with a given dielectric constant of the liquid medium,[7] understanding the dielectric properties of dipolar liquids under confinement is important to comprehend these processes.[29] The dielectric properties of liquids exhibit profound changes at the interface and upon confinement.[14, 21-22, 30-32] This occurs because of severely quenched fluctuations. Although the dielectric properties of bulk dipolar liquids are well understood,[33-36] there appears to be a limited number of studies devoted to understanding the dielectric behaviour of confined liquids. Furthermore, even the value of static dielectric constant inside nanocavities remains unclear.

One can write the Hamiltonian ($H$) and total interaction potential energy ($U$) for confined liquid systems in the following fashion [Eq.(1)].[37]

$$H = H_{liquid}^{(0)} + U(\boldsymbol{r}, \boldsymbol{R})$$
$$U(\boldsymbol{r}, \boldsymbol{R}) = \sum_{i, j(>i)} u_{ij}(\boldsymbol{r}_{ij}) + \sum_{i,k} u_{ik}(\boldsymbol{R}_{ik}) \qquad (1)$$

Here, $H_{liquid}^{(0)}$ denotes the kinetic energy of the fluid, $u_{ij}(\boldsymbol{r}_{ij})$ represents the intermolecular interactions in the liquid, and $u_{ik}(\boldsymbol{R}_{ik})$ represents the interaction of the liquid atoms/molecules with the surface atoms. $i$ and $j$ are the indices of liquid molecules. $k$ is the index of surface atoms. $\boldsymbol{r}$ and $\boldsymbol{R}$ denote the separation vectors between molecular/atomic centres. In practice, one models $u_{ij}(\boldsymbol{r}_{ij})$ as the sum of electrostatic, dipolar and Lennard-Jones interactions. However, $u_{ik}(\boldsymbol{R}_{ik})$ can be modeled in several different ways in order to characterize different surfaces.

Simulation-based studies provide microscopic insights. Nevertheless, the finite size of the systems and periodic boundary condition restrict the contributions from long wavelength modes. In a simulation study, Chandra *et al.* showed that the static dielectric constant ($\varepsilon_o$) of water decreases by approximately 50% inside a cavity of diameter 12.2 Å. The calculated values converge to the bulk by 24.4 Å diameter.[22] They claimed that their results remain consistent with two water models (SSD and SPC/E) studied. The same group studied the dielectric properties of model Stockmayer fluid and found similar trends.[38] On the other hand, White and co-workers reported a static dielectric constant of approximately 5 inside a smooth spherical cavity of 13.5 Å filled with SPC water.[32] However, they chose the dielectric constant of the wall as 5 in order to mimic the glass/mica surface. Recently, in an experimental study, Geim and co-workers have determined the value of the out-of-plane static dielectric constant as ~2 for an interfacial layer of water confined between two graphene sheets.[21]

In fact, the effects of geometric confinement and surface-liquid interactions have remained a subject of discussion for quite some time.[39-42] In the case of water, both the effects might be more complex. This is because of the extended hydrogen bond network (HBN). In order to minimize the free energy of the system, water molecules strive to maintain the HBN. This is often termed as the *principle of minimal frustration*.[43-45] However, water exhibits several anomalies and uniqueness. Hence, in this paper, we study another model dipolar liquid (Stockmayer fluid) to establish some general perspectives.

We raise and aim to answer the following questions. (i) How does the static dielectric constant scale with the size of the nanocavity? (ii) To what extent does the dielectric relaxation get modified in confinement? (iii) What is/are the microscopic origin(s) of faster collective orientational relaxation? (iv) How does the surface-liquid interaction affect the structure and dynamics of dipolar liquids?

The rest of the paper is organized as follows. In **section II**, we discuss the theoretical formalisms. In **section III**, we provide the derivation of Berendsen's equation from first principles and discuss its applicability in the nanoworld. **Section IV** contains the simulation details and parameters. In **section V**, we report the calculated values of static dielectric constant and its dependence on the size of the nanosphere. In **section VI**, we report and analyze the anomalous collective and single particle orientational relaxations under nanoconfinement. In **section VII**, we provide the angle distributions of molecular dipoles that reveal the altered structure of confined water molecules. **Section VIII** contains solvation dynamics studies and in **section IX** we discuss the origin of anomalous dielectric relaxation with the help of an Ising-Model based treatment. Finally, we summarise and discuss the future directions with some general conclusions in **section X**.

## II. Theoretical Formalism

According to the macroscopic theory of dielectrics,[46] evaluation of static dielectric constant $\varepsilon_0$ requires determination of the ratio of polarisation ($\boldsymbol{P}$) to the Maxwell field ($\boldsymbol{E}$) [Eq. (2)].

$$\varepsilon_0 = 1 + \frac{4\pi \boldsymbol{P}}{\boldsymbol{E}} = 1 + \left(\frac{4\pi}{V}\right)\frac{\boldsymbol{M}}{\boldsymbol{E}} \qquad (2)$$

$\boldsymbol{P}$ is calculated as the total dipole moment per unit volume ($\boldsymbol{M}/V$). This, in turn, requires the use of Kubo's linear response theory (LRT)[47]. However, one needs to account for a specific geometry and boundary conditions. For spherical samples, *Clausius-Mossotti relation provides the only exact expression for the static dielectric constant, $\varepsilon_0$* [Eq. (3)].

$$\frac{\varepsilon_0 - 1}{\varepsilon_0 + 2} = \frac{4\pi}{3V}\alpha \qquad (3)$$

Here, $V$ denotes the volume of the spherical sample and $\alpha$ represents the macroscopic polarisability. One can derive Eq. (3) starting from Maxwell's equations.[46, 48-49] However, this assumes that the surrounding medium is non-polarisable (vacuum), that is, $\varepsilon_{surr} = 1$. It is often convenient to use the frequency dependent counterpart of Eq. (3).

$$\frac{\varepsilon(\omega) - 1}{\varepsilon(\omega) + 2} = \frac{4\pi}{3V}\alpha(\omega) \qquad (4)$$

We shall work mostly with Eq. (3) in this study. By using the linear response theory (LRT) of Kubo[47], the frequency dependent polarizability $\alpha(\omega)$ [Eq. (4)] can be expressed as a Fourier transform of the *after effect function*, $b(t)$ [Eq. (5)].

$$\alpha(\omega) = -\int\limits_0^\infty dt\, e^{i\omega t}\frac{db(t)}{dt} \qquad (5)$$

One can relate $b(t)$ to the total dipole moment autocorrelation function, again by the application of LRT, as follows

$$b(t) = \frac{1}{3k_B T} <\boldsymbol{M}(0).\boldsymbol{M}(t)> . \qquad (6)$$

Equations (4), (5) and (6) lead to the expression for static dielectric constant ($\varepsilon_0$) of a spherical sample of volume $V$, suspended in vacuum ($\varepsilon_{surr} = 1$). Hence, the Clausius-Mossotti Eq. (3) for $\varepsilon_0$ becomes,

$$\frac{\varepsilon_0 - 1}{\varepsilon_0 + 2} = \frac{4\pi}{9Vk_B T}\left\langle M^2 \right\rangle_S \qquad (7)$$

In Eq. (7) the subscript '$S$' denotes spherical sample. In principle, $\left\langle M \right\rangle$, that is, the time-averaged total dipole moment, of the liquid confined inside a sphere should be zero. This ensures a proper sampling of the phase space.[37] However, in practice, we often find that for a finite system, and in a short time average $\left\langle M \right\rangle \neq 0$. This happens primarily because of the finite trajectory length. Hence, we replace $\left\langle M^2 \right\rangle_S$ by $\left\langle \delta M^2 \right\rangle_S$ in Eq. (7).

While Eq. (7) is exact, one needs a different expression to discuss the dielectric constant of a virtual sphere embedded in a spherical cavity. Such an expression was derived by Berendsen *et al.* to obtain static dielectric constant of a concentric spherical domain of radius $r_0$ inside a larger spherical domain of radius $R_c$ [Eq. (8)].[22, 50]

$$\frac{\left\langle M(r_0)^2 \right\rangle}{3k_s T r_0^3} = \frac{(\varepsilon - 1)}{9\varepsilon(\varepsilon + 2)}\left[(\varepsilon + 2)(2\varepsilon + 1) - 2(\varepsilon - 1)^2\left(\frac{r_0}{R_c}\right)^3\right] \qquad (8)$$

Here, $M(r_0)$ stands for the total dipole moment of the virtual sphere of radius $r_0$ (where $r_0 \leq R_c$). However, Berendsen's approach assumes that the dielectric constant of the smaller sphere (radius $\leq r_0$) to be the same as the outer shell ($r_0 <$ radius $\leq R_c$). Eq. (8) reduces to the Clausius-Mossotti relation for $r_0 = R_c$ and to the well-known Onsager-Kirkwood relation for $R_c \rightarrow \infty$.[35] One obtains the total dipole moment fluctuation from simulations. In **section III**, we detail the derivation of Eq. (8) with a discussion on its applicability.

On the other hand, one can derive the expression of static dielectric constant for a rectangular box of liquid with periodic boundaries, starting from Eq.(2). The expression [Eq.(9)] again assumes LRT for polarisability ($P$) and acquires the following form.

$$\varepsilon = 1 + \frac{4\pi}{3Vk_B T}\left\langle (\delta M)^2 \right\rangle \qquad (9)$$

Use of periodic boundary condition in Eq. (9) introduces approximations. The value of the dielectric constant, calculated from Eq.(9), approaches the bulk value even for small-sized systems as it contains the effect of periodic boundaries. On the other hand, while the Clausius-Mossotti equation is exact, the calculation of the static dielectric constant of the medium inside the sphere requires the creation of a surface and requires the use of several surface-liquid interactions.

Another delicate issue is the determination of the effective volume. As detailed in the subsequent sections, the Clausius-Mossotti relation shows a strong sensitivity to the volume $V$. Volume is determined by the nature of the surface-liquid interactions. If the surface is described by a collection of soft-repulsive spheres, then the interaction excludes a portion of the volume. This poses a problem of far-reaching consequences. In the usual applications of statistical mechanics, we probe the volume $V$ from outside. We do not account for the solute-solvent interactions. Of course, the goal of statistical mechanics is to consider the limit of $V \rightarrow \infty$ in order to recover the thermodynamic properties correctly. However, in the nanoscopic systems, that limit becomes inapplicable.[51]

In some earlier studies, a separate independent calculation of the dielectric constant of different sized cavities was not carried out. Also, the dipole moment cross-correlations were not evaluated. Instead, the dielectric constant of the liquid in the largest cavity was assumed to be same as the bulk value. Moreover, the volume of the cavity was not estimated systematically. Therefore, the values obtained remain doubtful. In **section V**, we address this issue in detail and prescribe an efficient solution.

## III. Static Dielectric Constant of Concentric Virtual Spheres: Berendsen's Equation

In this section, we derive Eq. (8) and discuss its applicability. We follow the method of Kirkwood.[35] We assume that the spherical sample of radius $R_c$ (with real boundary) is suspended in vacuum. Now we consider another smaller sphere of radius $r_0$ (with an imaginary boundary) that is concentric with the former. The outer spherical shell is treated as continuum dielectric with the same dielectric constant as that of the inner cavity. We also consider a fixed dipole $\boldsymbol{\mu}^*$ at the center (**Figure 1**).

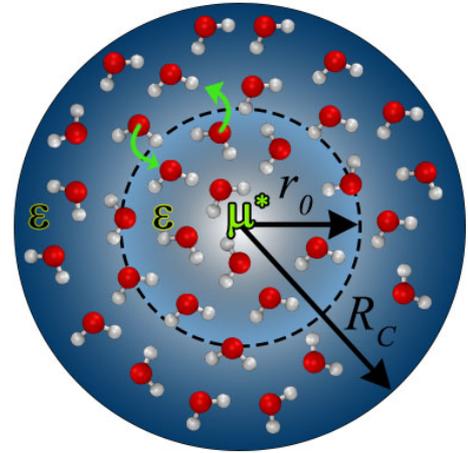

**Figure 1.** We show a schematic diagram that represents Berendsen's scheme and its underlying assumptions. The smaller sphere of radius $r_0$ is assumed to be enclosed by an imaginary spherical boundary (black dashed circle) that is concentric with the larger sphere. However, the larger sphere of radius $R_c$ possess a real boundary. The whole system is suspended in vacuum. There exists a fixed dipole ($\boldsymbol{\mu}^*$) at the centre. This particular formalism also assumes that the static dielectric constant ($\varepsilon$) inside the imaginary surface is the same as that of the outer shell.

Hence, one can divide the total dipole moment ($M$) of the spherical sample of volume $V$, produced by the fixed dipole into two parts as follows [Eq. (10)],

$$\boldsymbol{M} = \boldsymbol{M}(r_0) + \int_{v_0}^{V} \mathbf{P}\,dv . \qquad (10)$$

Here, $M(r_0)$ is the total dipole moment of the smaller sphere (with volume $v_0$) and $\mathbf{P}$ is the polarisation of the spherical shell that surrounds the smaller sphere. $\mathbf{P}$ is given by the following expression [Eq. (11)],[46, 49]

$$\mathbf{P} = -\frac{(\varepsilon - 1)}{4\pi} \nabla \psi_i \qquad (11)$$

where, $\psi_i$ is the electrostatic potential of the interior region. We now replace $\mathbf{P}$ in Eq. (10) and convert the volume integral to two surface integrals by the use of Green's theorem.[49]

$$\boldsymbol{M} = \boldsymbol{M}(r_0) - \frac{\varepsilon - 1}{4\pi}\left[\int_S \psi_i d\hat{s} - \int_{s_0} \psi_i d\hat{s}\right]. \qquad (12)$$

In Eq. (12), $S$ and $s_0$ respectively are the enclosing spherical surfaces of regions $V$ and $v_0$. We note that, because of symmetry considerations, one must consider the z-projection of the unit sphere surface element $(ds_z)$.[49] Hence, $ds_z = ds \cos\theta = \left(r^2 \sin\theta d\theta d\phi\right)\cos\theta$.

One can write the form of the interior potential ($\psi_i$) and the outside free space potential ($\psi_e$) according to the general solutions of the Laplace equation that uses Legendre polynomial expansion [Eq. (13)].[35, 48-49]

$$\psi_i(r_0 < r < R_c) = \sum_{n=1}^{\infty} A_n r^{-(n+1)} P_n(\cos\theta)$$
$$+ \sum_{n=1}^{\infty} B_n r^n P_n(\cos\theta) \qquad (13)$$
$$\psi_e(r > R_c) = \sum_{n=1}^{\infty} C_n r^{-(n+1)} P_n(\cos\theta)$$

Next, we impose two boundary conditions in order to maintain the continuity of the potential functions across the boundary.[46, 48-49]

$$-\varepsilon \left.\frac{\partial \psi_i}{\partial r}\right|_{r=R_c} = -\left.\frac{\partial \psi_e}{\partial r}\right|_{r=R_c}$$
$$-\frac{1}{R_c}\left.\frac{\partial \psi_i}{\partial \theta}\right|_{r=R_c} = -\frac{1}{R_c}\left.\frac{\partial \psi_e}{\partial \theta}\right|_{r=R_c} \qquad (14)$$

Solutions of Eqs. (14) yield the relations among the coefficients ($A_n$, $B_n$ and $C_n$) as depicted in Eqs. (15).

$$C_n = \frac{2n+1}{n\varepsilon + n + 1} A_n$$
$$B_n = \frac{n+1}{\varepsilon R^{2n+1}} \frac{\varepsilon - 1}{n\varepsilon + n + 1} A_n \qquad (15)$$

For dipolar systems, higher-order terms (n>1) do not contribute.[48] Additionally, as the static dielectric constant of the outside free space is unity, the coefficient $C_1$ becomes the net dipole moment of the whole sphere ($\mathbf{M}$). On the other hand, similar argument reveals $A_1$ is $\boldsymbol{\mu}^*$. Hence, one can simplify Eqs. (15) as follows.

$$\mathbf{M} = \frac{3}{\varepsilon + 2} \boldsymbol{\mu}^*$$
$$B_1 = \frac{2(\varepsilon - 1)}{\varepsilon R^2(\varepsilon + 2)} \boldsymbol{\mu}^* \qquad (16)$$

Now, we use the coefficients from Eq. (16) in Eq. (13) to evaluate the integrals in Eq. (12). The final expressions after evaluating the integrals are the following,

$$\int_S \psi_i d\hat{s} = \left(\frac{4\pi}{\varepsilon + 2}\right) \boldsymbol{\mu}^*$$
$$\int_{s_0} \psi_i d\hat{s} = \frac{8\pi(\varepsilon - 1)}{3\varepsilon(\varepsilon + 2)}\left(\frac{r_0}{R_c}\right)^3 + \frac{4\pi}{3}\left(\frac{\boldsymbol{\mu}^*}{\varepsilon}\right) \qquad (17)$$

Hence, Eq. (12) acquires the following form [Eq. (18)],

$$\frac{4\pi}{3v_0}\frac{\left\langle M(r_0)^2\right\rangle}{3k_B T} = \frac{\varepsilon - 1}{\varepsilon + 2}\left[1 + \frac{2}{9}\frac{(\varepsilon - 1)^2}{\varepsilon}\left\{1 - \left(\frac{r_0}{R_c}\right)^3\right\}\right] \qquad (18)$$

Eq. (18) is Berendsen's equation which is the same as Eq. (8) after minor rearrangements.[50] This equation can also be derived in a different way as demonstrated by Bossis.[52]

However, we raise certain concerns regarding the applicability of Eq. (18) for nanoscopic systems – (i) This method assumes that the dielectric constant of the inner sphere is the same as that of the outer shell. This is indeed true when both $r_0$ and $R_c$ are sufficiently large.[35] Nevertheless, for systems and subsystems that consist of ~$10^2$-$10^3$ water molecules, this assumption becomes invalid. (ii) Berendsen's derivation constructs an imaginary boundary through which molecules can escape and enter. This gives rise to significant density fluctuation. We note that the fluctuations in number density become negligible only if the sample size is large. (iii) The total dipole moments of the two regions, that is, the inner sphere of radius $r_0$ and the enclosing outer spherical shell, are correlated. This correlation becomes stronger as we decrease the sample size. Derivation of Eq. (18) does not consider this cross-correlation. Hence, application of Berendsen's equation on nano-confined fluid would provide unreal and erroneous values of static dielectric constant. Earlier studies that employed this equation to report dielectric properties of nanoconfined fluid remain doubtful.

## IV. Simulation Details

We perform atomistic molecular dynamics simulations of SPC/E water molecules and Stockmayer fluid. Below we provide the details of simulations and parameters for water and Stockmayer fluid separately.

**(a) Simulation of SPC/E water**: We consider three different liquid-surface potentials in order to model the surfaces- (i) atomistic wall with LJ-12,6 potential, (ii) virtual walls with LJ-9,3 potential [Eq. (19)] and (iii) virtual wall with LJ-10,4,3 potential [Eq. (20)].

$$U_{LJ}^{9-3}(r) = \frac{2}{3}\pi\rho_s\sigma_{sl}^3\varepsilon_{sl}\left[\frac{2}{15}\left(\frac{\sigma_{sl}}{r}\right)^9 - \left(\frac{\sigma_{sl}}{r}\right)^3\right] \quad (19)$$

$$U_{LJ}^{10-4-3}(r) = 2\pi\rho_s\sigma_{sl}^3\varepsilon_{sl}\left[\frac{2}{5}\left(\frac{\sigma_{sl}}{r}\right)^{10} - \left(\frac{\sigma_{sl}}{r}\right)^4 - \frac{\sqrt{2}\sigma_{sl}^3}{3\left(r+0.43\sigma_{sl}\right)^3}\right] \quad (20)$$

For atomistic walls, we choose the wall atom density ($\rho_s$) as that of the graphene sheet; $\sigma_s = 0.34$

nm and $\varepsilon_s = 0.09$ kcal/mol. We obtain the parameters for surface-water interactions as, $\sigma_{sl} = \left\{\left(\sigma_{surface} + \sigma_{liquid}\right)/2\right\}$ and $\varepsilon_{sl} = \sqrt{\varepsilon_{surface}\varepsilon_{liquid}}$; with $\sigma_{liquid} = 0.316\,nm$ and $\varepsilon_{liquid} = 0.155\,Kcal/mol$. We model the walls as non-polarisable and uncharged. In the case of (ii) and (iii), we evaluate the surface-water interaction energy between water molecules and the closest point on the virtual sphere.

In the case of atomistic walls, we simulate six spherical nano-cavities of radii ($R_c$) = 1.0 nm ($N_{wat} = 70$), 1.5 nm ($N_{wat} = 306$), 2.0 nm ($N_{wat} = 812$), 2.5 nm ($N_{wat} = 1,695$), 3.0 nm ($N_{wat} = 3,059$) and 4.0 nm ($N_{wat} = 7,691$). We perform simulations with virtual walls for five different cavities of radii 1.17 nm ($N_{wat} = 140$), 1.67 nm ($N_{wat} = 638$), 2.17 nm ($N_{wat} = 1,116$), 3.17 nm ($N_{wat} = 3,765$) and 4.17 nm ($N_{wat} = 10,064$). We separately simulate 4,142 SPC/E water molecules in a 5nm cubic box with periodic boundary conditions (PBC) to calculate the required bulk properties for comparison. We use NVT (T = 300 K) ensemble with Nose-Hoover chain thermostat ($\tau = 0.21\,ps^{-1}$). For bulk simulations, we use particle mesh Ewald to obtain long-range electrostatics with an FFT grid spacing of 0.16 nm.

**(b) Simulation of Stockmayer fluid**: We simulate nanocavities of radii 1.17 nm (N=87), 2.17 nm (N=680), 3.17 nm (N=2,300), and 4.17 nm (N=5,453). The LJ parameters are as follows, $\sigma_{liquid} = 0.34$ nm and $\varepsilon_{liquid} = 0.23$ kcal/mol. We perform the simulations with $\rho^* = 0.8$, $\mu^* = 1.0 (= 0.81 D)$ and $T^* = 1.0$. Other simulation protocols remain similar to that of water. We model the surface using LJ-9,3 potential [Eq. (19)] with the same parameters. For the bulk system, we simulate 500 particles with PBC.

We carry out the cavity simulations without PBC and without cut-off for either long range or short range interactions. We perform the simulations for 10 ns and analyze the last 8 ns. We use LAMMPS[53] and GROMACS[54] to produce the MD trajectories. We employ in-house codes written in FORTRAN and MATLAB for analyses. We use VMD[55] for visualization purposes.

# V. Static dielectric constant of confined dipolar liquid

The dielectric constant of solvent governs electrostatic screening. This, in turn, can affect the encounter probability of solute molecules. Low static dielectric constant also results in slow solvation. Hence, evaluation and understanding of dielectric constant become a topic of paramount importance. In this section, we report static dielectric constants ($\varepsilon_0$) of water and Stockmayer fluid in spherical nanoscopic confinements. We employ the *exact* relation which is the Clausius-Mossotti equation [Eq.(7)] and use the effective/accessible volumes as described later in this section. We plot the calculated values in **Figure 2.**

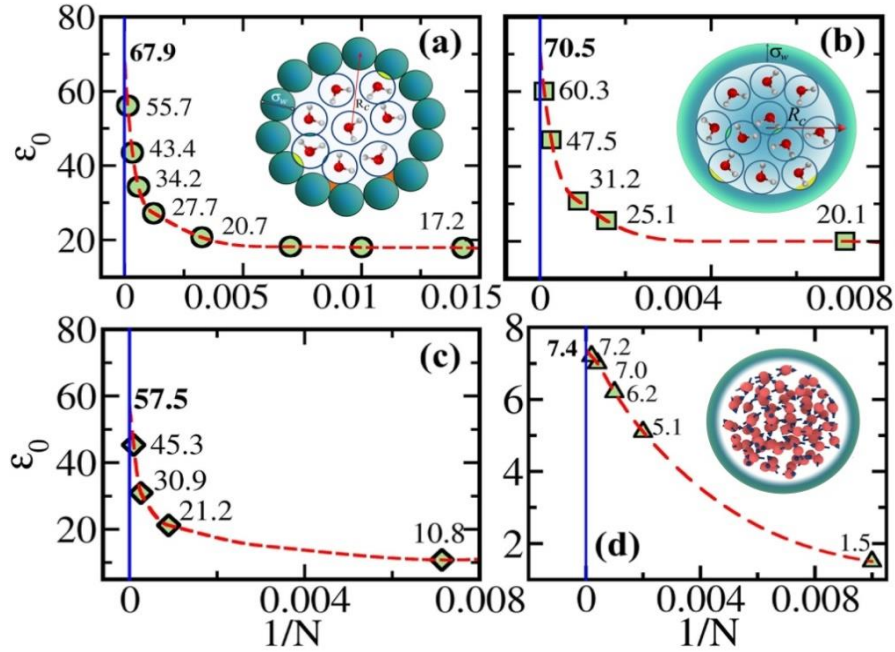

**Figure 2. Static dielectric constant ($\varepsilon_0$) against the inverse of the number of molecules (1/N) for aqueous nanocavities with (a) LJ-12,6 atomistic walls, (b) LJ-9,3 walls, (c) LJ-10,4,3 walls and (d) Stockmayer fluid with LJ-9,3 wall. The convergence for water is extremely slow. Extrapolations using a cubic polynomial provide limiting values of 67.9, 70.5 and 57.5 respectively in the thermodynamic limit. However, the convergence for Stockmayer fluid is remarkably fast. (Insets) we show schematic two-dimensional cross-sections of the nanocavities. Penetration of water molecules to the soft-spheres (yellow regions) and inaccessibility of certain regions (orange regions) inside the cavity invoke errors in the volume calculation.**

It is clear that the static dielectric constant of spherically nanoconfined water shows a strong size dependence and slow convergence to the bulk value. However, the dielectric constant of Stockmayer fluid reaches nearly the bulk value by ~3 nm (**Figure 2d**).

In order to obtain the value of $\varepsilon_0$ in the thermodynamic limit (N→∞), we plot $\varepsilon_0$ against 1/N. We extrapolate the data by the use of cubic spline polynomials. Extrapolations provide good agreements with the bulk value– 67.9 for atomistic LJ-12,6 wall (**Figure 2a**) and 70.5 for virtual LJ-9,3 wall (**Figure 2b**). However, LJ-10,4,3 wall provides a much lower value (~58) (**Figure 2c**) compared to the bulk $\varepsilon_0$ (~68) with periodic boundaries upon extrapolation. We shall discuss the origin of such low dielectric constant in subsequent sections.

***Volume calculation and sensitivity of $\varepsilon_0$ to volume.***
Determination of the accessible volume becomes crucial to obtain the value of $\varepsilon_0$. We rearrange Eq. (7) to obtain the following expression [Eq. (21)]

$$\varepsilon_0 = \frac{8\pi\beta\left\langle\delta M^2\right\rangle + 9V}{9V - 4\pi\beta\left\langle\delta M^2\right\rangle}. \quad (21)$$

The denominator on the right-hand side of Eq. (21) becomes zero if $9V = 4\pi\beta\left\langle\delta M^2\right\rangle$. Hence, $\varepsilon_0$ diverges (**Figure 4**). In the case of periodic or macroscopic systems, volume calculation becomes trivial and error free. However, it remains nontrivial in nanoconfined systems, especially when the liquid is surrounded by a soft-repulsive wall.

One of the ways is to determine the volume *post facto*. That is, we estimate the accessible volume ($V_{eff}$) from computer simulations in the following fashion. We first obtain radial population distributions of oxygen atoms with respect to the center of the sphere (**Figure 3**). We observe how close water molecules reach the wall atoms. We integrate over the density distribution and normalize to the total number of particles ($N$).

$$\frac{1}{N}\int_0^{R_{eff}} dr\, 4\pi r^2 \rho(r) = 1 \quad (22)$$

We use Eq. (22) to numerically obtain $R_{eff}$ ($< R_C$) from the distributions shown in **Figure 3**. This provides a measure of accessible volume, $V_{eff} = (4/3)\pi R_{eff}^3$. We use this effective volume in Clausius-Mossotti equation.

In order to demonstrate the sensitivity of $\varepsilon_0$ to $V$, we plot $\varepsilon_0$ against $(1/R)$ for the same $\left\langle\delta M^2\right\rangle$ in **Figure 4(a)**, **4(b)** and **4(c)**. We vary the effective radius from $R_c - \left(\sigma_W/2\right)$ to $R_c$ for $R_c=1$ nm system.

It is clear from **Figure 4** that a minute change in the calculation of the effective radius can lead to a noticeable change in the value of $\varepsilon_0$. For example, in $R_C=1$nm system, if one changes $R_{eff}$ from 0.93 nm to 0.90 nm, the value of $\varepsilon_0$ changes from 17.2 to 46.1. Hence, careful determination of $V_{eff}$ becomes crucial. However, we do not observe such a strong dependence on volume for confined Stockmayer fluid.

In an earlier study, Chandra and Bagchi showed a similar divergence like behaviour of wavenumber dependent dielectric function [$\varepsilon(k)$] in dipolar liquids [**Figure 4(d)**].[56]

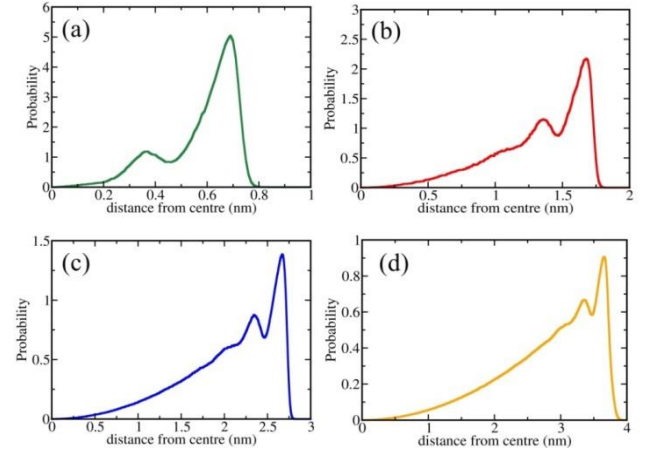

**Figure 3. Plots show normalized population distributions of oxygen atoms of water with respect to the center of the nanocavities of radii- (a) 1.0 nm, (b) 2.0 nm, (c) 3.0 nm, and (d) 4.0 nm. Data in these plots correspond to water molecules that are confined inside atomistic LJ-12,6 walls. These distributions provide a measure of inaccessible regions inside the cavity. We use effective radius calculated from the distributions to obtain static dielectric constant from the Clausius-Mossotti equation. We obtain similar distributions for virtual walls and Stockmayer fluid.**

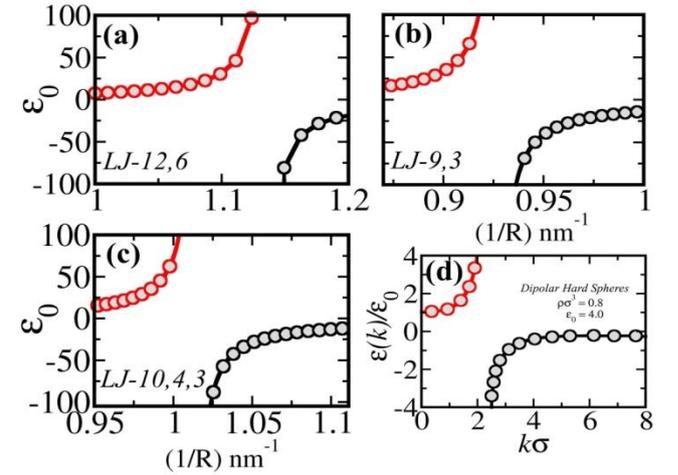

**Figure 4. $\varepsilon_0$ of water against the inverse of effective radius for $R_c=1$nm aqueous cavity. Surfaces are described by– (a) LJ-12,6, (b) LJ-9,3, and (c) LJ-10,4,3 potentials. The sensitivity of $\varepsilon_0$ to $R_{eff}$ is quite strong. After the divergent like behaviour $\varepsilon_0$ becomes negative. (d) Wave vector ($k$) dependent dielectric function calculated for dipolar hard sphere liquid by employing mean spherical approximation (MSA) approach, shows similar divergence like behaviour.[56]**

## VI. Anomalous polarisation relaxation and single particle rotation

Rotational motions of solvent molecules play an important role in the solvation process of a solute. At the early stages of aqueous solvation dynamics, libration and single particle rotation contribute approximately 60-80%.[57-58] Here we calculate collective orientational correlations (that is, the total dipole moment autocorrelation, $C_M(t)$) and single particle rotational correlations ($C_1(t)$) for confined water and Stockmayer fluid [Eq.(23)].

$$C_M(t) = \langle \boldsymbol{M}(0).\boldsymbol{M}(t) \rangle = \left\langle \sum_i \boldsymbol{\mu}_i(0). \sum_i \boldsymbol{\mu}_i(t). \right\rangle$$

$$C_1(t) = \frac{1}{N} \sum_{i=1}^{N} \left\langle P_1\left(\hat{\boldsymbol{\eta}}_o^i . \hat{\boldsymbol{\eta}}_t^i\right) \right\rangle; \text{ where } P_1(x) = \cos x \quad (23)$$

Here, $\hat{\boldsymbol{\eta}}_t^i$ is the unit vector along one of the O—H bonds of $i^{th}$ water molecule at time $t$. We average over all the water molecules and obtain particle averaged decay.

Surprisingly, we observe twenty times faster dipole moment relaxation in the case of nanoconfined water as compared to the bulk. This observation remains independent of the chosen surface-liquid interaction and size of the cavity. Bulk dielectric relaxation exhibits single exponential decay with a time constant of ~10 ps (SPC/E water at 300K). Experimentally it is found to be ~8.3 ps.[59] On the other hand, dielectric relaxations of cavity water molecules exhibit bi-exponential decay with time constants ~30 fs (40%) and ~700 fs (60%) (**Figure 5a**). We also observe a slightly faster single particle rotational relaxation. We find two timescales- (i) in the ~200-600 fs regime (10-20%), and (ii) in the ~3-4 ps regime (80-90%). However, the timescales are comparable to that in the bulk (**Figure 5b**). In **Figure 5c** and **5d**, we show the collective and single particle rotational relaxations for confined Stockmayer fluid. We find that the total dipole moment relaxation is approximately four times faster than the bulk relaxation. However, the single particle rotation is slower than the bulk, although with comparable timescales. We provide the bi-exponential fitting parameters and average timescales in **Table 1**.

**Table 1.** (a) Bi-exponential fitting parameters for the total dipole moment autocorrelations inside aqueous nanocavities and in the bulk. The relaxation timescales for cavity water are in the order of femtoseconds. Whereas, the <M(0).M(t)> relaxation in the bulk SPC/E water is found to be single exponential and associated with ~10 ps timescale. (b) Multi-exponential fitting parameters for particle averaged first rank rotational time correlation functions of confined water molecules inside nanospheres and in the bulk. We obtain two distinct timescales- one in the order of 200-600 fs and another in the order of ~3-4 ps. These timescales of confined water are, however, comparable to the bulk.

| Cavity Radius | (a) Collective orientational relaxation | | | | | (b) Single particle orientational relaxation | | | | |
|---|---|---|---|---|---|---|---|---|---|---|
| | $a_1$ | $\tau_1$ (ps) | $a_2$ | $\tau_2$ (ps) | $<\tau>$ (ps) | $a_1$ | $\tau_1$ (ps) | $a_2$ | $\tau_2$ (ps) | $<\tau>$ (ps) |
| 1 nm | 0.42 | 0.03 | 0.58 | 0.73 | 0.44 | 0.25 | 0.60 | 0.75 | 4.3 | 3.4 |
| 2 nm | 0.41 | 0.04 | 0.59 | 0.78 | 0.48 | 0.19 | 0.34 | 0.81 | 4.6 | 3.8 |
| 3 nm | 0.37 | 0.02 | 0.63 | 0.59 | 0.38 | 0.16 | 0.28 | 0.84 | 4.7 | 4.0 |
| 4 nm | 0.40 | 0.03 | 0.60 | 0.70 | 0.43 | 0.14 | 0.26 | 0.86 | 4.7 | 4.1 |
| Bulk | 1.00 | 10.4 | --- | --- | 10.4 | 0.13 | 0.21 | 0.87 | 4.9 | 4.3 |

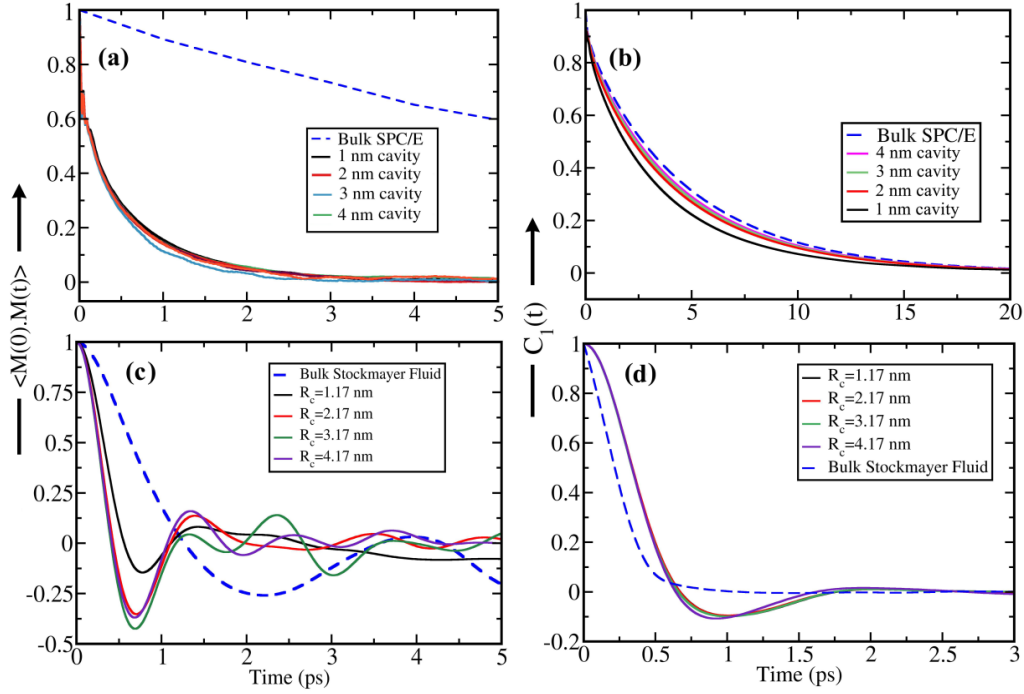

**Figure 5.** (a) Normalised total dipole moment autocorrelation function (collective orientational relaxation) of water for spherical nano-cavities of different radii. Surprisingly, the relaxation of the total dipole moment relaxation in confinement is approximately twenty times faster compared to that in the bulk (blue dashed line). (b) Normalised and particle averaged first rank orientational time correlation function for spherically confined water molecules. The confined water molecules show a slightly faster decay than that in the bulk while the opposite is expected. The distinction is prominent, especially for $R_c = 1$ nm. However, as we increase the size of the nano-cavity the decay converges to the bulk response (blue dashed line). (c) Total dipole moment autocorrelation function of stockmayer fluid inside spherical nano-cavities. We observe that the relaxation of the total dipole moment in confinement is approximately four times faster compared to that in the bulk (blue dashed line). (d) Normalized and particle averaged first rank orientational time correlation function for spherically confined Stockmayer fluid. However, unlike water, the confined particles show a slower decay than that in the bulk, however, with comparable timescales. If we increase the size of the nano-cavity the decay patterns do not approach the bulk response (blue dashed line).

We provide an explanation of such anomalous ultrafast decays by means of the interplay among self- and cross-correlations of different regions inside the cavity. We divide the system into two, four and eight equal regions to obtain region-specific collective dipole moments. The time trajectories of such region specific dipole moments reveal the correlation length in the system. In **Figure 6**, we plot the time trajectories of angles made by the total dipole moment vectors of two hemispheres with two of the Cartesian axes for $R_c=4$ nm aqueous and Stockmayer fluid system with atomistic walls. We observe signatures of strong anti-correlation for two of the direction cosines associated with a Pearson's correlation coefficient ($\rho_{ij}$) ~ -0.85, where, $i$ and $j$ are the indices of two different regions (here, two hemispheres). However along the third axis it shows weak correlation ($\rho_{ij}$ ~0.1, graph not shown). We observe similar trends for $R_c=1$ nm, 2 nm and 3 nm systems as well. However, for Stockmayer fulid, such anti-correlations are rather weak with $\rho_{ij}$ ~-0.3-0.4 (**Figure 6c** and **6d**). We observe such anti-correlations in direction cosines for concentric spheres of smaller radii inside the nanocavity.

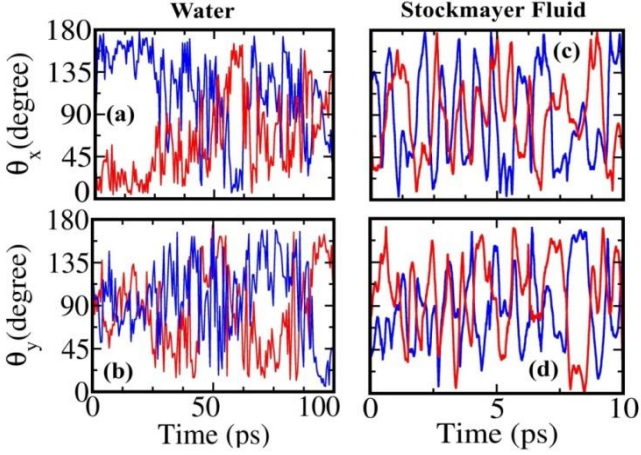

Figure 6. Time evolution of angles created by the total dipole moment vectors of two hemispheres with (a) X-axis and (b) Y-axis for $R_c$=4 nm aqueous system with atomistic LJ-12,6 walls. (c) and (d) show similar plots for $R_c$=4 nm Stockmayer fluid system with LJ-9,3 wall. In Stockmayer fluid, the fluctuations are short lived. These plots show strong anti-correlated dipole flips that result in enormous cancellations.

We next describe the total dipole moment time-correlation function in terms of sub-ensembles. This reveals the timescales of anti-correlations. If we divide the spherical sample into '$m$' equal sub-ensembles, there are $m$ number of self-terms ( $\langle M_i(0).M_i(t) \rangle$ ) and $^mP_2 = \frac{m(m-1)}{2}$ number of cross-terms $\langle M_i(0).M_j(t) \rangle$, where $i$ and $j$ denote the indices that represent different grids [Eq. (24)].

$$\langle M(0).M(t) \rangle = \sum_{i=1}^{m} \langle M_i(0).M_i(t) \rangle + \sum_{i,j=1}^{m} \langle M_i(0).M_j(t) \rangle \quad (24)$$

Figure 7 (for $m$=8) shows the presence of anti-correlation among the coarse-grained dipole moments of eight grids. In the case of $m$=2 and 4, we observe similar behaviour. Although the amplitudes of self-terms are ~4-10 times larger than that of the cross terms, the total negative contribution that arises from the anti-correlated cross terms makes the resultant decay ultrafast (red dashed curves in Figure 7).

Furthermore, the power spectrum of DMTCF for nanoconfined water exhibit bimodal *1/f* noise (Figure 8). We attribute the deviation from bulk exponent

(~0.9) to the surface effect and heterogeneous dynamics inside the nanocavities.

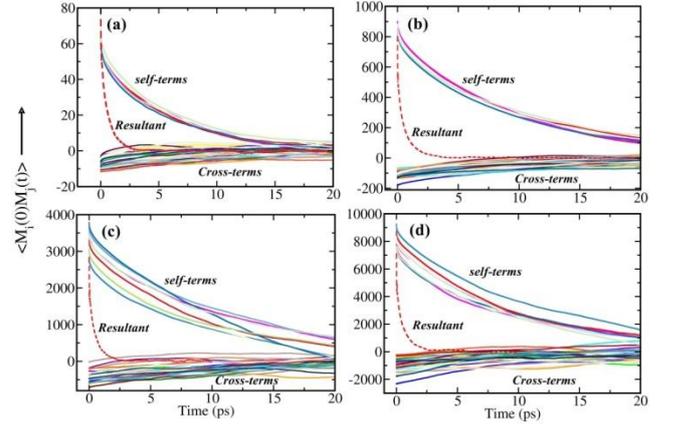

Figure 7. The plots show self- and cross-dipole moment correlations among eight grids inside aqueous nano-cavities of radius (a) $R_c$=1 nm, (b) $R_c$=2 nm, (c) $R_c$=3 nm, and (d) $R_c$=4 nm. The amplitudes of self-terms are ~4-10 times higher than that of the cross-terms. However, there are eight self-terms and 56 cross-terms for each system that construct the total <M(0).M(t)> for a cavity. As a result, the negative contributions from anti-correlated cross terms predominate at longer times. This makes the net relaxation (red dashed lines) ultrafast.

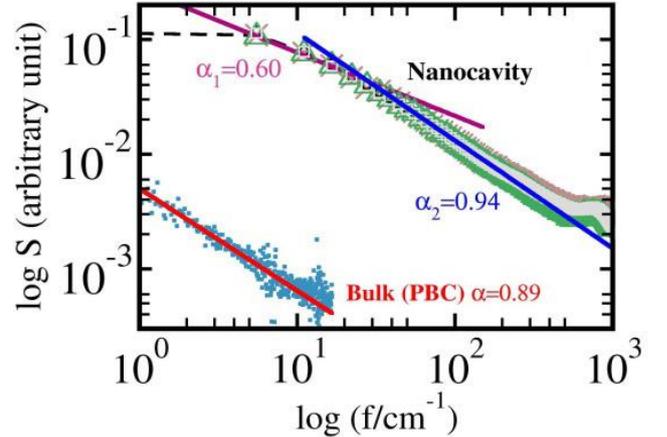

Figure 8. Bimodal *1/f* character of the power spectrum of dipole moment fluctuation under confinement and in the bulk. Bulk power spectrum decays with a single power law exponent approximately equal to 0.9. On the other hand, the power spectrum of nanoconfined water (all sizes) exhibit two different exponents- one on the lower (0.60) and another on the higher side (0.94). This deviation and bimodality can be attributed to the surface effects and the heterogeneous dynamics of liquid water under nanoconfinement.

In order to rationalize the results shown in Figure 6 and Figure 7, we inquire the *Kirkwood g-factor* ( $g_K$ ) [Eq.(25)] for nanocavities and bulk. $g_K$

reveals information on the microscopic ordering of molecular dipoles. $g_K$ becomes unity if the relative orientations of dipoles are random.

$$g_K = \left\langle M^2 \right\rangle \Big/ N\mu^2 \qquad (25)$$

$g_K$ for confined water varies from 0.15 to 0.21, whereas in the bulk (with PBC) $g_K$ of SPC/E water is

~3.6 at 300 K. In an earlier simulation study of intermediate-sized water clusters, Ohmine *et al.* reported similar low values of $g_K$.[14] Under confinement, Stockmayer fluid also exhibits an approximately four-fold decrease in the values of $g_K$. In **Table 2,** we report the numerical values.

**Table 2. Values of *Kirkwood g-factor* of water and Stockmayer fluid for different kinds of surface-liquid interactions. We find that $g_K$ reduces substantially in confinement compared to the bulk value. In the case of water, it becomes almost twenty times smaller than the bulk. However, for Stockmayer fluid, it becomes approximately four times smaller than the bulk. The calculated values of $g_K$ remain independent of the size of the nanocavity. As the average timescale of dielectric relaxation is proportional to $g_K$, it provides a quantitative explanation of the faster than bulk decay of <M(0).M(t)>.**

| *Water* | | | | | *Stockmayer fluid* | |
|---|---|---|---|---|---|---|
| **Radius** | $g_K$ **(12,6)** | **Radius** | $g_K$ **(9,3)** | $g_K$ **(10,4,3)** | **Radius** | $g_K$ **(9,3)** |
| 1 nm | 0.21 | 1.17 nm | 0.20 | 0.16 | 1.17 nm | 0.17 |
| 2 nm | 0.18 | 2.17 nm | 0.17 | 0.15 | 2.17 nm | 0.69 |
| 3 nm | 0.17 | 3.37 nm | 0.16 | 0.16 | 3.17 nm | 0.68 |
| 4 nm | 0.17 | 4.47 nm | 0.17 | 0.15 | 4.17 nm | 0.66 |
| Bulk = 3.64 | | | | | Bulk = 2.03 | |

The timescale of collective orientational relaxation ($\tau_M$) is related to $g_K$ and single particle rotational correlation timescale ($\tau_S$) by the following relation [Eq. (26)]

$$\tau_M = \frac{g_K}{g_K^D(0)}\tau_S. \qquad (26)$$

Here, $g_K^D(\omega)$ is the frequency dependent dynamic Kirkwood g-factor. This can be expressed as,

$$g_K^D(\omega) = \frac{\sum\limits_{i,j}^{N}\int\limits_0^\infty dt\, e^{-i\omega t}\left\langle \dot{\mu}_i(0)\dot{\mu}_j(t)\right\rangle}{\sum\limits_i^{N}\int\limits_0^\infty dt\, e^{-i\omega t}\left\langle \dot{\mu}_i(0)\dot{\mu}_i(t)\right\rangle} \qquad (27)$$

We find that $g_K$ reduces substantially in confinement compared to the bulk value. Such reductions are independent of the size of cavities. In the case of water, it becomes almost twenty times smaller than the bulk. However, for Stockmayer fluid, it becomes approximately four times smaller than the bulk. This indicates that the collective alignment of microscopic dipoles is destructive inside nanocavities resulting in *enormous cancellations among correlations*. On the other hand, in periodic bulk systems, the microscopic dipoles align constructively. We obtain $g_K^D(0)$ in between 1.4 to 1.8 inside aqueous nanocavities and 1.5 for bulk SPC/E water. The deviation from bulk, in this case, is not significant. Hence, from Eq.(26), $\tau_M$ becomes approximately proportional to $g_K$. This explains the faster collective relaxation inside the cavity.

## VII. Surface orientation and tetrahedral network:

We perform layer-wise analyses (each layer is taken to be 5Å thick) to observe the differences in relative orientations as we approach the center of the sphere. We plot the distributions of angles formed between O—H bonds of water and the surface normal (**Figure 9a**). We observe a distinct peak around ~90° for the outermost layer of water (**Figure 9b-9e**). This advocates the preservation of certain preferred orientations near the surface. Similar observations have been made by Ruiz-Barragan *et al.* from *ab initio* simulations of water confined inside graphene slit-pores.[60] The water molecules follow the *principle of minimal frustration* often used to describe protein folding and spin-glass transitions.[43-45] In this case, it occurs through the maximization of hydrogen bonds in order to minimize the free energy of the system. In an earlier simulation study, Banerjee *et al.* reported similar observations for two dimensional Mercedes-Benz model confined between two hydrophobic plates.[61] However, in the case of Stockmayer fluid we cannot make any distinction between surface layers and interior dipoles in terms of such distributions (**Figure 10**).

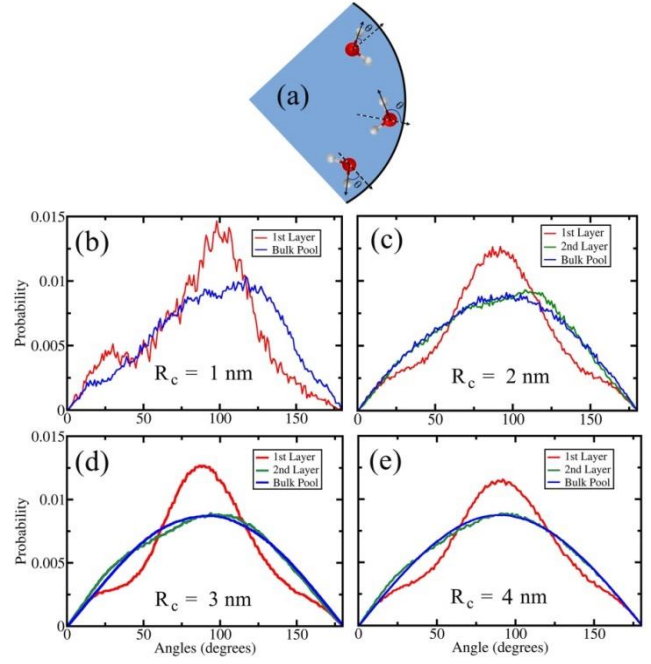

**Figure 9. (a) A schematic representation of the orientation of water molecules relative to the surface normal. We plot distributions of the angles ($\theta$) shown in this figure for different layers of water inside nano-cavities of radii (b) $R_c = 1$ nm, (c) $R_c = 2$ nm, (d) $R_c = 3$ nm, and (d) $R_c = 4$ nm. In all these cases, the surface layer shows distinct characteristic. The angle distributions for the surface layer show a distinct peak around ~90°. This depicts the preservation of certain preferred orientations. In order to minimize the free energy of the system, water molecules strive to maintain the hydrogen bond network. This is termed as the *principle of minimal frustration in protein folding and spin-glass transition literature*. However, as we approach the center, the layers and the central bulk pool show similar distributions. (*The above results are obtained for water molecules trapped inside atomistic LJ-12,6 walls. We obtain similar plots for other systems.*)**

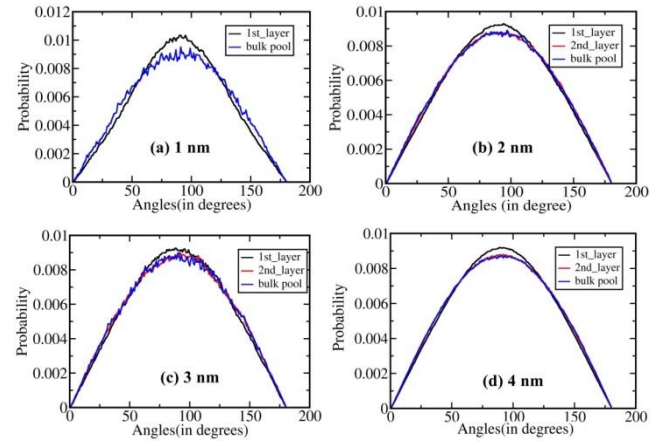

**Figure 10. Distribution of relative orientation of Stockmayer fluid dipoles with the surface normal of the enclosing LJ-9,3 wall. Unlike water, the surface layer exhibits no distinctness in terms of preferred orientations.**

We plot the distribution of O-O-O angles in order to observe alterations in the tetrahedral network because of confinement. In this calculation, we reject the contribution of the outermost layer because that layer is not surrounded by other water molecules uniformly from all sides. We note that, consideration of the outermost shell can introduce artefacts. We observe two distinct peaks – (i) a broad peak centered at 110°, and (ii) a smaller peak centered at 60°. The distributions overlap with each other (**Figure 11**). Hence, we conclude that the spatial network structure of water remains unperturbed inside nanocavities.

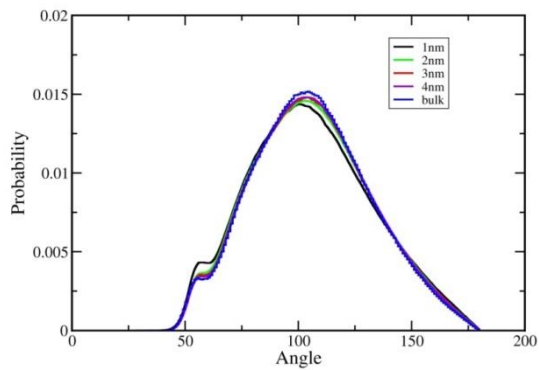

**Figure 11. Distribution of O-O-O angles of water inside nano-cavities of different radii. The distribution shows a small peak near 60° and a broad peak centered at 110°. We obtain the distributions for the nano-cavities without considering the outermost layer of water molecules that remain in contact with the wall atoms. This is because those water molecules are not surrounded by other water molecules uniformly from all directions. Nanocavities larger than 1nm radius exhibit almost similar O-O-O angle distribution. This indicates the indifference of spatial structure inside nanocavities. (The above results are obtained for water molecules trapped inside atomistic LJ walls. We obtain similar plots for other systems.)**

## VIII. Aqueous solvation dynamics inside nanocavity

Dipolar solvation dynamics is one of the most important aspects of chemical dynamics. It provides a measure of how fast can perturbed charge distribution of a solute get stabilized by solvent reorientation. According to the continuum model description, the solvation energy relaxation timescale ($\tau_L$) is related to Debye relaxation timescale ($\tau_D$) as, $\tau_L = \left(\varepsilon_\infty / \varepsilon_0\right)\tau_D$.[57, 62] Here, $\varepsilon_\infty$ and $\varepsilon_0$ are the infinite frequency and zero frequency dielectric constant of the dipolar continuum respectively.

In our study, we place a frozen water molecule at the center of the cavity (or box). We use this as the probe. We calculate the energy autocorrelation function as $C_s(t) = \langle\delta E(0)\delta E(t)\rangle / \langle\ \delta E(0)^2\rangle$. In the regime of linear response approximation, $C_s(t)$ and the non-equilibrium stokes shift response function, $S(t)$ become equivalent.[57] We fit the resultant decay using multi-exponential functions. Except for $R_c = 1$ nm cavity, the solvation energy relaxations show a similar decay pattern as that in the bulk (**Figure 12**).

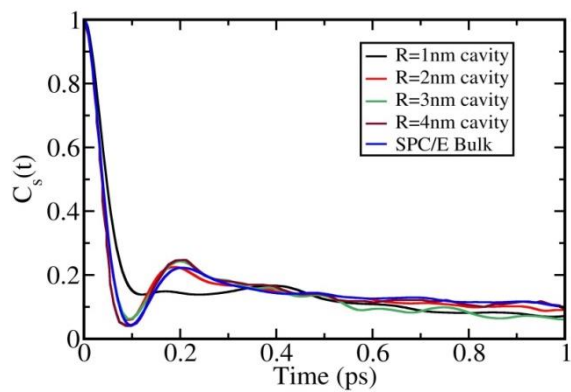

**Figure 12. Solvation energy time correlation function (TCF) for a frozen water molecule (probe) situated at the center of the nanocavity. In the bulk, we follow the same procedure for a periodic cubic box. We calculate the solvation energy as the sum of electrostatic and LJ interactions of the frozen water molecules with all the other water molecules in the system. Solvation shows ultrafast characteristics with a ~30-40 fs component that contributes 80% of the decay. The rest 20% lies in the ~1-2 ps range. The nature of solvation TCF converges that of the bulk by $R_c$=2 nm cavity.**

We find that solvation is ultrafast. Almost 80% of the initial decay occurs in the ~30-40 fs timescale. This is because of single particle rotation and libration of the surrounding water molecules. We find another timescale in the ~1-2 ps time regime (with ~20% contribution). The relatively slower timescale arises because of collective hydrogen bond reorientations of surrounding water molecules.

## IX. An Ising model based theoretical explanation of the *faster than Bulk Relaxation*

The *faster than bulk* rotational and dielectric relaxation inside the cavity can be attributed to surface effects. Biswas *et al.* developed a theoretical description to explain this phenomenon.[1] Their approach follows kinetic Ising model[63] and Glauber dynamics.[64] One can discuss this model in terms of a one dimensional Ising chain (that is along a diameter of the sphere) with a Hamiltonian $H = -J\sum_{<ij>}\sigma_i\sigma_j$ and then extend this approach for a two-dimensional *spin on a ring* model.[27] One of the assumptions of this model is that the diametrically opposite water molecules possess opposite relative orientation. However, other water molecules can freely rotate (**Figure 13**).

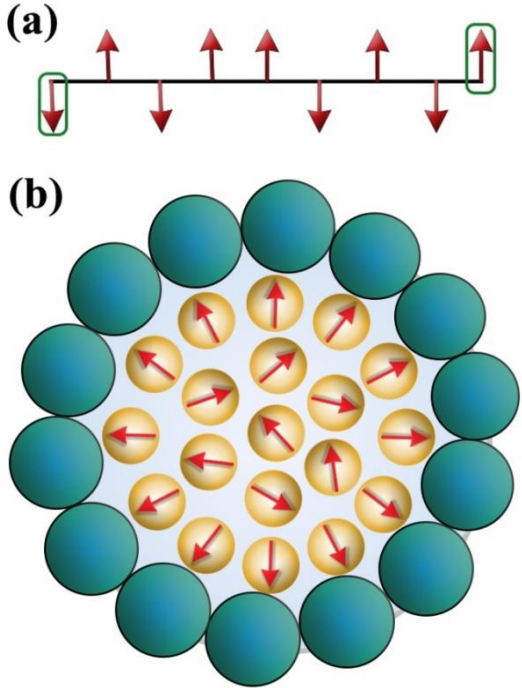

**Figure 13. A schematic description of the chosen model- (a) One-dimensional Ising chain with two fixed and opposite spins at the two ends (green box). These fixed spins represent surface effects. The spins in the middle (bulk pool) can freely fluctuate. (b) Two-dimensional spin on a ring model system. The dipoles (water molecules or Stockmayer fluid) close to the surface preserve their spins because of preferred orientations and owing to the principle of minimal frustration. If we look through any one of the diameters, we recover the one-dimensional model.**

Let the probability of the spin state that consists of $N$ number of spins $(\sigma_1, \sigma_2, ..., \sigma_N)$ at time $t$ be $p(\sigma_1, \sigma_2, ..., \sigma_N, t)$. One can write the Glauber master equation as [Eq. (28)],

$$\frac{d}{dt}p(\sigma_1, \sigma_2, ..., \sigma_N, t) = -\sum_j w_j(\sigma_{j-1}, \sigma_j, \sigma_{j+1})p(\sigma_1, \sigma_2, ..., \sigma_j, ..., \sigma_N, t)$$
$$+ \sum_j w_j(\sigma_{j-1}, -\sigma_j, \sigma_{j+1})p(\sigma_1, \sigma_2, ..., -\sigma_j, ..., \sigma_N, t)$$

$$(28)$$

Here, the transition probability, $w_j(\sigma_{j-1}, \sigma_j, \sigma_{j+1}) = \frac{1}{2}\alpha\left[1 - \frac{1}{2}\gamma\sigma_j(\sigma_{j-1} + \sigma_{j+1})\right]$ and $\gamma = \tanh(2J\beta)$. $\alpha/2$ is the rate of spin transition per unit time. Now, one defines a stochastic dynamical variable $q(t)$ to obtain the expectation value of $j^{th}$ spin [Eq. (29)].

$$q_j(t) = \langle\sigma_j(t)\rangle = \sum\sigma_j(t)p(\sigma_1, \sigma_2, ..., \sigma_N, t) \qquad (29)$$

By the use of the definitions of $q_j(t)$ and $w_j$ in Eq. (28), one arrives at the following equation of motion for $j^{th}$ spin [Eq. (30)]

$$\frac{dq_j(t)}{dt} = -\alpha q_j(t) + \frac{\alpha\gamma}{2}\left[q_{j-1}(t) + q_{j+1}(t)\right]. \qquad (30)$$

Solutions of Eq. (30) exist with various boundary conditions. Eq. (31) describes a continuum model description of Eq. (30). The continuum model description is more suited for spins that reside away from the surface.

$$\frac{\partial q(x,t)}{\partial t} = -\alpha(1-\gamma)q(x,t) + \frac{a^2\alpha\gamma}{2}\frac{\partial^2 q(x,t)}{\partial x^2}. \qquad (31)$$

Here, $a$ denotes the lattice spacing. One can easily solve Eq. (31) at the low-temperature limit ($\gamma \to 1$) and boundary conditions, $q(0,t) = 1$, $q(L,t) = -1$ and $q(x,0) = f(x)$. We provide the analytical expression for $q(x,t)$ in Eq. (32).

$$q(x,t) = 1 - 2\frac{x}{L} - \frac{2}{\pi}\sum_{n=1}^{\infty}\left[\frac{\cos n\pi + 1}{n}\sin\frac{n\pi x}{L}\right.$$
$$\left. + \frac{2}{L}\sin\frac{n\pi x}{L}\int_0^L f(x')\sin\frac{n\pi x'}{L}dx'\right]e^{-\left(\frac{a^2\alpha n^2\pi^2 t}{2L^2}\right)}$$

$$(32)$$

The temporal relaxation of $q(x,t)$ becomes faster as $x \to (L/2)$ from both the ends. We can define the total dipole moment of the system, at any given time step $t$ as

$$M(t) = \int_0^L dx \, q(x,t) \, . \qquad (33)$$

Hence, we write the total dipole autocorrelation in terms of the autocorrelation of the stochastic variable $q(x,t)$ as,

$$\langle \boldsymbol{M}(0)\boldsymbol{M}(t) \rangle = \int_0^L dx \int_0^L dx' \langle q(x,0)\,q(x',t) \rangle \quad (34)$$

This model shows that the spins closer to the center relax at a faster rate. This faster relaxation of $\langle q(0)q(t) \rangle_x$ gets reflected in the collective relaxation of moment-moment time correlation as indeed observed from our simulations. However, the spins near the surface relax on a slower time scale, reflecting the effect of the surface.

## X. Summary and Conclusion

In conventional discussions of liquid state properties, we usually do not need to discuss the effects of the surface. Also, the surface effects are expected to diminish beyond a distnce of few molecular diameters. However, the situation can be different fron dipoar liquids, especially for water where not only long range dipolar interactions but also hydrogen bond network can get altred in a significant way.

The dielectric constant of a liquid is a collective property, determined by the long wavelength orientational correlations in the system.[11-12, 56, 65] Because of the long wavelength nature of the orientational correlations in dipolar liquids, the dielectric constant is rather strongly dependent on size and shape. Use of periodic boundary conditions in most simulations thus introduces an approximation which needs to be tackled carefully. The approach via the Clausius-Mossotti equation is exact but one has to deal with a slow convergence. For water, this convergence is particularly slow due to the extensive hydrogen bond network of water. As discussed in detail in this work, the problem becomes more acute in the nanoworld.

In this paper, we report a comprehensive study of the dielectric properties of water and Stockmayer fluid confined to spherical nanocavities. Such a study by varying the size of the nanocavity and water-surface interactions was not carried out before. The study gave rise to many new results, some of which are rather interesting on science ground. Below, we summarise the key outcomes.

i.      We find a substantial reduction in the static dielectric constant ($\varepsilon_0$) of nanoconfined water. The convergence toward the bulk value is slow. However, $\varepsilon_0$ of Stockmayer fluid shows a weaker dependence on the size of the nanocavity.

ii.      We derive Berendsen's equation [Eq. (8) or (18)] by following Onsager-Kirkwood formalism.[35] This assumes the static dielectric constant of the concentric inner sphere is the same as that of the enclosing spherical shell which is, in turn, assumed to be a continuum. We find that such an assumption becomes invalid in nano-dimensions. This is because of frequent particle exchange through the imaginary inner boundary and also because of the presence of strong spatiotemporal correlations among the dipole moments of different regions inside the cavity. *Berendsen's equation is only asymptotically valid.*

iii.      We show that the Clausius-Mossotti equation is rather sensitive to the volume of the system. *In nanoscopic world volume is defined by intermolecular interactions, unlike the macroscopic description of volume, that is prescribed from outside.* When the enclosing surface is modeled as soft spheres, effective volume calculation is subject to errors. We show that a small error in $V_{eff}$ leads to substantial changes in $\varepsilon_0$ for nanoconfined water. However, Stockmayer fluid does not exhibit noticeable changes in $\varepsilon_0$ with $V_{eff}$.

We have employed an effective method here that needs further refinement. Our method correctly reproduces the bulk value on extrapolation. In some sense, *Figure 2 is quite remarkable*. For water (SPC/E model), the value of the static dielectric constant is already within 20% at $R_C$=4 nm and within 10% at $R_C$=4.6 nm. Our way of extrapolation provides a true measure of the static dielectric constant at the thermodynamic limit as it does not contain the artefact imposed by PBC.

iv.      We encounter a surprising result that, *the total dipole autocorrelations ($C_M(t)$) decay approximately twenty times faster for nanoconfined water as* compared to the bulk response. $C_M(t)$ of nanoconfined water also exhibit a bimodality in the power spectrum. $C_M(t)$ decays approximately four times faster for confined Stockmayer fluid systems. Furthermore, the timescales of relaxation do not change with the increasing size of the cavity, within the sizes considered. We explain the anomalous fast relaxation in terms of *substantially low values of Kirkwood g-factor* and also in terms of *anti-correlated local dipole moments* of different regions inside the cavity.

v. Solvation dynamics, single particle rotational correlations and tetrahedrality show much faster convergence to the bulk with increasing cavity size.

vi. Nature of the surface-liquid interaction affects the values of $\varepsilon_0$ but does not alter the general trends. We have confirmed this claim by using five different surface-water interactions.

vii. The above results demonstrate that the anomalies arise solely because of geometric confinement.

In our study, the surfaces and the surrounding medium are taken as non-polarisable materials. Clausius-Mossotti equation demands the surrounding medium to be non-polarizable ($\varepsilon_{surr} = 1$). However, the surface should in practice be polarizable. One of the ways to introduce polarisability is by considering the wall atoms as Drude oscillators. The other approach would be to perform *ab initio* simulations. Also, the effect of the shape of confinement and the origin of dielectric anisotropy remain relatively less explored. We have planned future works in this direction.

## Acknowledgment

We thank Professor R. N. Zare for early collaboration. We also thank Professor Dominik Marx for pointing out the works of Netz and Hansen. BB thanks Sir J. C. Bose fellowship, SM thanks UGC and DST for providing financial support, and SA thanks IISc for the scholarship.

## References

1. Biswas, R.; Bagchi, B. A kinetic Ising model study of dynamical correlations in confined fluids: Emergence of both fast and slow time scales. *J. Chem. Phys.* **2010**, *133*, 084509.

2. Bagchi, B. *Water in Biological and Chemical Processes: From Structure and Dynamics to Function*. Cambridge University Press: 2013.

3. Bhattacharyya, K.; Bagchi, B. Slow dynamics of constrained water in complex geometries. *J. Phys. Chem. A* **2000**, *104*, 10603-10613.

4. Laage, D.; Elsaesser, T.; Hynes, J. T. Water Dynamics in the Hydration Shells of Biomolecules. *Chem. Rev.* **2017**, *117*, 10694–10725.

5. Bhattacharyya, K. Solvation dynamics and proton transfer in supramolecular assemblies. *Acc. chem. Res.* **2003**, *36*, 95-101.

6. Fleming, G. R.; Wolynes, P. G. Chemical dynamics in solution. *Physics Today* **1990**, *43*, 36-43.

7. Bagchi, B. *Molecular relaxation in liquids*. Oxford University Press, USA: New York, 2012.

8. Pollock, E.; Alder, B. Static dielectric properties of stockmayer fluids. *Physica A: Statistical Mechanics and its Applications* **1980**, *102*, 1-21.

9. Maroncelli, M.; Fleming, G. R. Computer simulation of the dynamics of aqueous solvation. *J. Chem. Phys.* **1988**, *89*, 5044-5069.

10. Perera, L.; Berkowitz, M. L. Dynamics of ion solvation in a Stockmayer fluid. *J. Chem. Phys.* **1992**, *96*, 3092-3101.

11. Chandra, A.; Bagchi, B. A molecular theory of collective orientational relaxation in pure and binary dipolar liquids. *J. Chem. Phys.* **1989**, *91*, 1829-1842.

12. Bagchi, B.; Chandra, A. Polarization relaxation, dielectric dispersion, and solvation dynamics in dense dipolar liquid. *J. Chem. Phys.* **1989**, *90*, 7338-7345.

13. Zhou, H. X.; Bagchi, B. Dielectric and orientational relaxation in a Brownian dipolar lattice. *J. Chem. Phys.* **1992**, *97*, 3610-3620.

14. Saito, S.; Ohmine, I. Dynamics and relaxation of an intermediate size water cluster (H2O) 108. *J. Chem. Phys.* **1994**, *101*, 6063-6075.

15. Munoz-Santiburcio, D.; Marx, D. Chemistry in nanoconfined water. *Chem. Sci.* **2017**, *8*, 3444-3452.

16. Muñoz-Santiburcio, D.; Marx, D. Nanoconfinement in Slit Pores Enhances Water Self-Dissociation. *Phys. Rev. Lett.* **2017**, *119*, 056002.

17. Wittekindt, C.; Marx, D. Water confined between sheets of mackinawite FeS minerals. *J. Chem. Phys.* **2012**, *137*, 054710.

18. Gekle, S.; Netz, R. R. Anisotropy in the dielectric spectrum of hydration water and its relation to water dynamics. *J. Chem. Phys.* **2012**, *137*, 104704.

19. Schlaich, A.; Knapp, E. W.; Netz, R. R. Water dielectric effects in planar confinement. *Phys. Rev. Lett.* **2016**, *117*, 048001.

20. Lee, J. K.; Banerjee, S.; Nam, H. G.; Zare, R. N. Acceleration of reaction in charged microdroplets. *Quart. Rev. Biophys.* **2015**, *48*, 437-444.

21.    Fumagalli, L.; Esfandiar, A.; Fabregas, R.; Hu, S.; Ares, P.; Janardanan, A.; Yang, Q.; Radha, m.; Taniguchi, T.; Watanabe, K.; Gomila, G.; Novoselov, K.; Geim, A. Anomalously low dielectric constant of confined water. *Science* **2018**, *360*, 1339-1342.

22.    Senapati, S.; Chandra, A. Dielectric constant of water confined in a nanocavity. *J. Phys. Chem. B* **2001**, *105*, 5106-5109.

23.    Mondal, S.; Acharya, S.; Biswas, R.; Bagchi, B.; Zare, R. N. Enhancement of reaction rate in small-sized droplets: A combined analytical and simulation study. *J. Chem. Phys.* **2018**, *148*, 244704.

24.    Girod, M.; Moyano, E.; Campbell, D. I.; Cooks, R. G. Accelerated bimolecular reactions in microdroplets studied by desorption electrospray ionization mass spectrometry. *Chem. Sci.* **2011**, *2*, 501-510.

25.    Nam, I.; Lee, J. K.; Nam, H. G.; Zare, R. N. Abiotic production of sugar phosphates and uridine ribonucleoside in aqueous microdroplets. *Proc. Natl. Acad. Sci. U.S.A.* **2017**, *114*, 12396-12400.

26.    Piletic, I. R.; Moilanen, D. E.; Spry, D.; Levinger, N. E.; Fayer, M. Testing the core/shell model of nanoconfined water in reverse micelles using linear and nonlinear IR spectroscopy. *J. Phys. Chem. A* **2006**, *110*, 4985-4999.

27.    Biswas, R.; Chakraborti, T.; Bagchi, B.; Ayappa, K. Non-monotonic, distance-dependent relaxation of water in reverse micelles: propagation of surface induced frustration along hydrogen bond networks. *J. Chem. Phys.* **2012**, *137*, 014515.

28.    Mondal, S.; Mukherjee, S.; Bagchi, B. Protein hydration dynamics: Much ado about nothing? *J. Phys. Chem. Lett.* **2017**, *8*, 4878-4882.

29.    Bagchi, B. Dynamics of solvation and charge transfer reactions in dipolar liquids. *Annu. Rev. Phys. Chem.* **1989**, *40*, 115-141.

30.    Blaak, R.; Hansen, J.-P. Dielectric response of a polar fluid trapped in a spherical nanocavity. *J. Chem. Phys.* **2006**, *124*, 144714.

31.    Ballenegger, V.; Hansen, J.-P. Dielectric permittivity profiles of confined polar fluids. *J. Chem. Phys.* **2005**, *122*, 114711.

32.    Zhang, L.; Davis, H. T.; Kroll, D.; White, H. S. Molecular dynamics simulations of water in a spherical cavity. *The Journal of Physical Chemistry* **1995**, *99*, 2878-2884.

33.    Debye, P. *Polar Molecules (Chemical Catalog, New York, 1929)*. 1954; p 84.

34.    Hubbard, J.; Onsager, L. Dielectric dispersion and dielectric friction in electrolyte solutions. I. *J. Chem. Phys.* **1977**, *67*, 4850-4857.

35.    Kirkwood, J. G. The dielectric polarization of polar liquids. *J. Chem. Phys.* **1939**, *7*, 911-919.

36.    Cole, K. S.; Cole, R. H. Dispersion and absorption in dielectrics I. Alternating current characteristics. *J. Chem. Phys.* **1941**, *9*, 341-351.

37.    Bagchi, B. *Statistical Mechanics for Chemistry and Materials Science* CRC Press: New York, 2018.

38.    Senapati, S.; Chandra, A. Molecular dynamics simulations of simple dipolar liquids in spherical cavity: Effects of confinement on structural, dielectric, and dynamical properties. *J. Chem. Phys.* **1999**, *111*, 1223-1230.

39.    Alba-Simionesco, C.; Coasne, B.; Dosseh, G.; Dudziak, G.; Gubbins, K.; Radhakrishnan, R.; Sliwinska-Bartkowiak, M. Effects of confinement on freezing and melting. *J. Phys. Cond. Mat.* **2006**, *18*, R15.

40.    Alcoutlabi, M.; McKenna, G. B. Effects of confinement on material behaviour at the nanometre size scale. *J. Phys. Cond. Mat.* **2005**, *17*, R461.

41.    Morineau, D.; Xia, Y.; Alba-Simionesco, C. Finite-size and surface effects on the glass transition of liquid toluene confined in cylindrical mesopores. *J. Chem. Phys.* **2002**, *117*, 8966-8972.

42.    Ping, G.; Yuan, J.; Vallieres, M.; Dong, H.; Sun, Z.; Wei, Y.; Li, F.; Lin, S. Effects of confinement on protein folding and protein stability. *J. Chem. Phys.* **2003**, *118*, 8042-8048.

43.    Wolynes, P. G.; Onuchic, J. N.; Thirumalai, D. Navigating the folding routes. *Science* **1995**, *267*, 1619-1621.

44.    Bryngelson, J. D.; Wolynes, P. G. Spin glasses and the statistical mechanics of protein folding. *Proc. Natl. Acad. Sci. U.S.A.* **1987**, *84*, 7524-7528.

45.    Onuchic, J. N.; Wolynes, P. G. Theory of protein folding. *Curr. Opin. Struct. Biol* **2004**, *14*, 70-75.

46.    Fröhlich, H. *Theory of dielectrics*. 1949.

47.    Kubo, R. In *Linear response theory of irreversible processes*, Statistical Mechanics of Equilibrium and Non-equilibrium, 1965; p 81.

48.    Jackson, J. D. *Classical electrodynamics*. John Wiley & Sons: 2012.

49.    Böttcher, C. J. F.; van Belle, O. C.; Bordewijk, P.; Rip, A. *Theory of electric polarization*. Elsevier Science Ltd: 1978; Vol. 2.

50.    Berendsen, D. J. Molecular Dynamics and Monte Carlo Calculations of Water. *CECAM report (unpublished)* **1972**, 29.

51.    Hill, T. L. *Thermodynamics of small systems*. Courier Corporation: 1994.

52.    Bossis, G. Molecular dynamics calculation of the dielectric constant without periodic boundary conditions. I. *Mol. Phys.* **1979**, *38*, 2023-2035.

53.    Plimpton, S.; Crozier, P.; Thompson, A. LAMMPS-large-scale atomic/molecular massively parallel simulator. *Sandia National Laboratories* **2007**, *18*, 43.

54.    Hess, B.; Kutzner, C.; Van Der Spoel, D.; Lindahl, E. GROMACS 4: algorithms for highly efficient, load-balanced, and scalable molecular simulation. *J. Chem. Theo. Comp.* **2008**, *4*, 435-447.

55.    Humphrey, W.; Dalke, A.; Schulten, K. VMD: visual molecular dynamics. *Journal of molecular graphics* **1996**, *14*, 33-38.

56.    Chandra, A.; Bagchi, B. Exotic dielectric behavior of polar liquids. *J. Chem. Phys.* **1989**, *91*, 3056-3060.

57. Bagchi, B.; Jana, B. Solvation dynamics in dipolar liquids. *Chem. Soc. Rev.* **2010,** *39*, 1936-1954.

58. Mondal, S.; Mukherjee, S.; Bagchi, B. Origin of diverse time scales in the protein hydration layer solvation dynamics: A Simulation Study. *J. Chem. Phys.* **2017,** *147*, 154901.

59. Buchner, R.; Barthel, J.; Stauber, J. The dielectric relaxation of water between 0 C and 35 C. *Chem. Phys. Lett.* **1999,** *306*, 57-63.

60. Ruiz-Barragan, S.; Muñoz-Santiburcio, D.; Marx, D. Nanoconfined Water within Graphene Slit Pores Adopt Distinct Confinement–Dependent Regimes. *J. Phys. Chem. Lett.* **2018**.

61. Banerjee, S.; Singh, R. S.; Bagchi, B. Orientational order as the origin of the long-range hydrophobic effect. *J. Chem. Phys.* **2015,** *142*, 04B602_1.

62. Castner Jr, E. W.; Fleming, G. R.; Bagchi, B.; Maroncelli, M. The dynamics of polar solvation: inhomogeneous dielectric continuum models. *J. Chem. Phys.* **1988,** *89*, 3519-3534.

63. Fredrickson, G. H.; Andersen, H. C. Kinetic ising model of the glass transition. *Phys. Rev. Lett.* **1984,** *53*, 1244.

64. Glauber, R. J. Time-dependent statistics of the Ising model. *J. Math. Phys.* **1963,** *4*, 294-307.

65. Bagchi, B.; Chandra, A. Macro–micro relations in dipolar orientational relaxation: An exactly solvable model of dielectric relaxation. *J. Chem. Phys.* **1990,** *93*, 1955-1958.